\documentclass{elsart}

\def\eeq{\end{equation}}
\def\beq{\begin{equation}}
\def\bea{\begin{eqnarray}}
\def\eea{\end{eqnarray}}

\usepackage{amssymb}
\usepackage{graphics}
\journal{Physica A}

\begin{document}
\begin{frontmatter}

\title{Dissipative maps at the chaos threshold: 
Numerical results for the single-site map}
\author{Ugur Tirnakli\thanksref{now}}
\thanks[now]{tirnakli@sci.ege.edu.tr}
\address{Department of Physics, Faculty of Science, Ege University, 35100 
Izmir, Turkey} 

\begin{abstract}
We numerically study, at the edge of chaos, the behaviour of the 
single-site map $x_{t+1}=x_t- x_t/(x^2_t +\gamma^2)$, where 
$\gamma$ is the map parameter. 
\end{abstract}
\end{frontmatter}

In recent years, a number of studies has addressed the nonlinear dynamical 
systems at the chaos 
threshold \cite{TPZ,costa,lyra,circle,latbar,ugur,ugur2,ugur3}. 
At this special point (and also at some others such as period doublings, 
tangent bifurcations points where the Lyapunov exponent vanishes), 
it is shown that the proper entropy appears to be the nonextensive 
form \cite{ts88} $S_q = k \left(1-\sum_{i=1}^W p_i^q\right)/(q-1)$ with 
a proper $q$ value (say $q^*$) for each map under consideration. 
Up to now, three different methods are available for the determination 
of the $q^*$ values. First method comes from the analysis of the 
sensitivity to initial conditions where a general sensitivity function 

\begin{equation}
\xi (t)=\left[ 1+(1-q)\lambda _qt\right] ^{1/(1-q)}\;\;\;\;\;\;
(q\in  R),
\end{equation}
can be introduced. Here, $\lambda_q$ is the generalized Lyapunov 
exponent and this form recovers the standard exponential one for 
$q=1$, but it generically exhibits a power-law behaviour. In fact, 
eq.(1) corresponds to the power-law growth of the upper bound of 
the sensitivity function and this upper bound allows us to determine 
the proper $q^*$ value of the studied map. 
Second method is based on the geometrical aspects of the attractor 
at the edge of chaos and uses the scaling behaviour of the 
multifractal singularity spectrum $f(\alpha)$ as   

\beq
\frac{1}{1-q^*} = \frac{1}{\alpha_{min}} - \frac{1}{\alpha_{max}}
\;\;\;\;\;\; (q^*<1)
\eeq
where $\alpha_{min}$ and $\alpha_{max}$ being the most concentrated 
and most rarefied regions of the measure on the attractor respectively. 
It is obvious that calculating these values from the $f(\alpha)$ 
curve, one can easily estimate the proper $q^*$ value \cite{lyra}.
Finally, third method, which was introduced very recently, deals 
with the entropy increase rates \cite{latbar,ugur}. It basically 
based upon the (ensemble version) of generalized Kolmogorov-Sinai 
entropy $K_q \equiv \lim_{t\rightarrow\infty}
\lim_{W\rightarrow\infty}\lim_{N\rightarrow\infty}
\left[S_q(t)\right]/t$ and it is conjectured that (i) a special 
$q^*$ value exists such that $K_q$ is finite for $q=q^*$, vanishes 
for $q>q^*$ and diverges for $q<q^*$; (ii) this $q^*$ value 
coincides with that coming from the other two methods explained above.

Now we are prepared to introduce the main purpose of the present 
work. Here, we shall try to verify once again these conjectures 
using the abovementioned three methods for the single-site map

\beq
x_{t+1}=x_t- \frac{x_t}{x^2_t +\gamma^2}\;\; ,
\eeq
where $\gamma$ is the map parameter \cite{single}. The shape of the 
map is illustrated in Fig.~1. This map has a route to chaos by period 
doubling with Feigenbaum scaling but it exhibits an interesting 
property that the bifurcation cascade splits into two independent 
ones, which both obey Feigenbaum scenario. This property is the 
generic peculiarity for the single-site map in contrast to the 
logistic map \cite{single}. On the other hand, these two independent 
cascades obey Feigenbaum scaling, which means that the Feigenbaum 
numbers of this map are the same as those of the logistic map, 
therefore, in that sense the single-site map shares the same 
universality class with the logistic map. Consequently, what we 
expect is that, at the chaos threshold, the proper $q^*$ value of 
this map would coincide with the value of the logistic case 
(namely, $q^*\simeq 0.24$). 
Our numerical results check and verify this expectation. 
Firstly, let us perform the first method to estimate the proper 
$q^*$ value of the single-site map at the onset of chaos 
(namely at $\gamma_c = 0.43400483...$). For the extremal point 
of the map ($x_0=0.3213...$), the sensitivity function exhibits 
a power-law divergence, $\xi(t)\propto t^{1/(1-q)}$ as shown in 
Fig.~2. Measuring the upper bound slope $1/(1-q)$, we determine 
the proper $q$ value as $q^*=0.24~$. Now, let us proceed with 
the second method. In Fig.~3, the $f(\alpha)$ curve is given, where 
the monotonic behaviour with respect to the number of iterations 
is evident. Since the correct values of $\alpha_{min}$ and 
$\alpha_{max}$ could be obtained in principle when the number of 
iterations goes to infinity, the extrapolation shown in the inset of 
Fig.~3 is used to estimate the correct values of $\alpha_{min}$ and 
$\alpha_{max}$. Using these values in the scaling relation Eq.(2), 
it is found that the proper $q$ value is again $q^*=0.24$. 
(We also verify using the extrapolation technique that the fractal 
dimension $d_f$ of this map coincides with that of the logistic map 
as expected). Finally, we shall use the third method for this map. 
To do this, we partition the phase space, $x\in (-1/2\gamma,1/2\gamma)$, 
into $W$ equal cells, then we choose one of them and select $N$ initial 
conditions all inside the chosen cell. As time evolves, these $N$ points 
spread within the phase space and finally this yields a set of 
probabilities. In the beginning of time, of course, $S_q(0)=0$, then 
it exhibits three successive regions. In the intermediate region, the 
entropy starts increasing and this is the region where the linear 
increase of the proper entropy is expected to emerge (for details 
see \cite{latbar,ugur}). In Fig.~4, we plotted the time evolution 
of the entropy for three values of $q$. (At this point, it should be 
mentioned that the procedure of averaging over the efficient cells as 
described in \cite{latbar,ugur} has also been performed here). 
In the intermediate region, it is seen that the linear increase of the 
entropy with time emerges for a special value of $q$. To support this 
picture, we fit the curve with the polynomial $S_q=A+Bt+Ct^2$ in the 
intermediate region $[t_1,t_2]$. Since the nonlinearity coefficient 
$R\equiv C(t_1+t_2)/B$ is a measure of the importance of the nonlinear 
term, it should vanish for a strickly linear fit. Therefore, inset 
of Fig.~4, we estimate the proper $q$ value as $q^*=0.234$, which is 
the value when $R$ is zero. 

Summing up, we study the behaviour of the single-site map at the 
chaos threshold. We numerically verify that the proper entropic index 
for this map is $q^*\simeq =0.24$. This constitutes another example 
of the one-dimensional dissipative maps which behave as $q\neq 1$ at 
the chaos threshold.

This work has been supported by Ege University Research Fund 
under the project number 2000 FEN 049.  

\newpage

{\bf Figure Captions}

{\bf Figure 1} - The shape of the single-site map with $\gamma =0.3~$.

{\bf Figure 2} - Log-log plot of the sensitivity function versus time.

{\bf Figure 3} - Approximate multifractal singularity spectrum of the 
critical attractor of the single-site map for increasingly large number 
of iterations, going from $N=64$ to $4096$. Inset: $\alpha_{min}$, $d_f$ 
and $\alpha_{max}$ from the singularity spectra obtained from distinct 
number of iterations. Extrapolated values of them are also given.

{\bf Figure 4} - Time evolution of $S_q(t)$ for three different values of $q$.
$W=10^5$ and $N=10^6$ is used. The interval characterizing the intermediate 
region is [9,21]. Inset: The nonlinearity coefficient $R$ versus $q$ 
(see text for details).

\newpage

\includegraphics{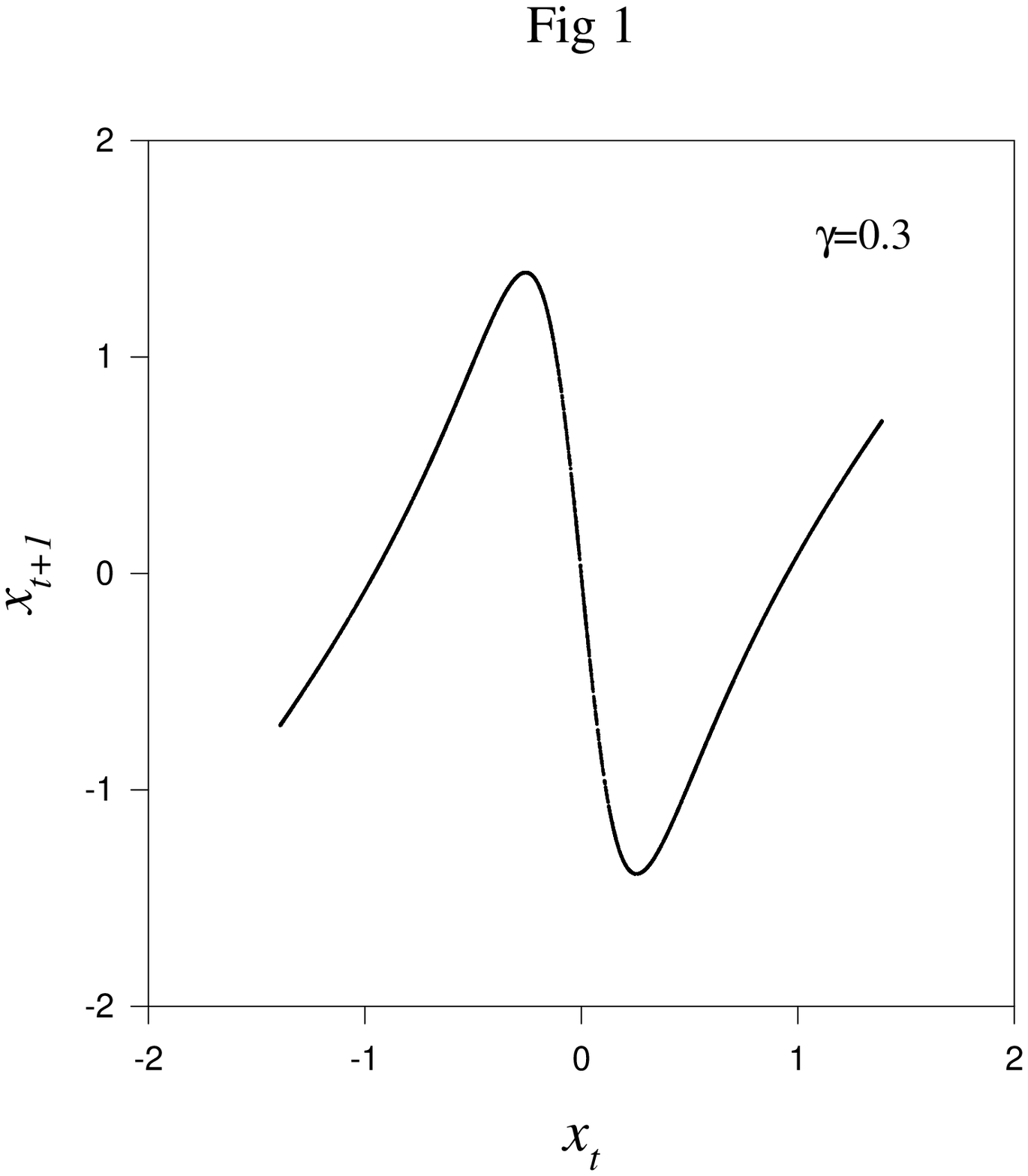}

\includegraphics{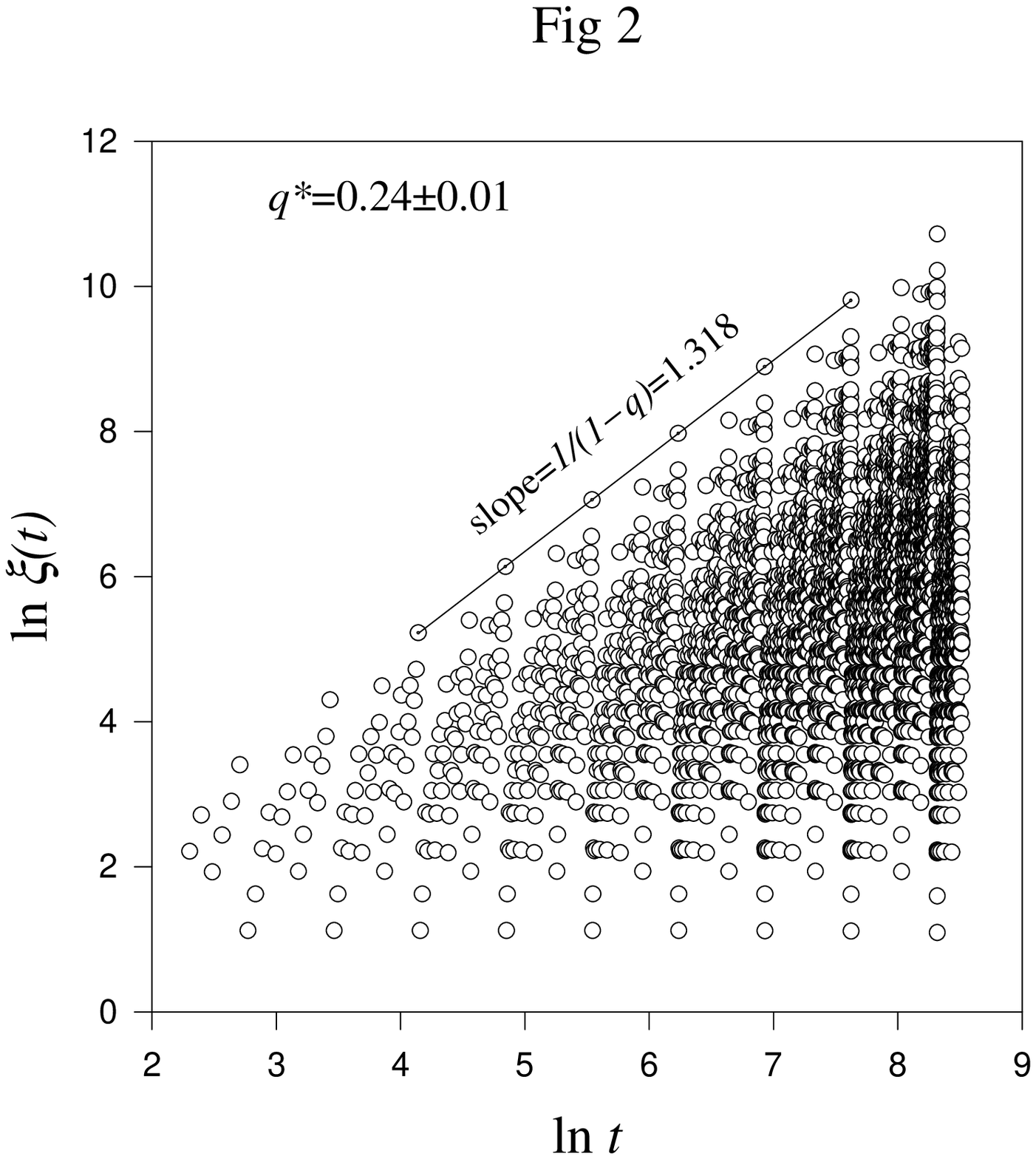}

\includegraphics{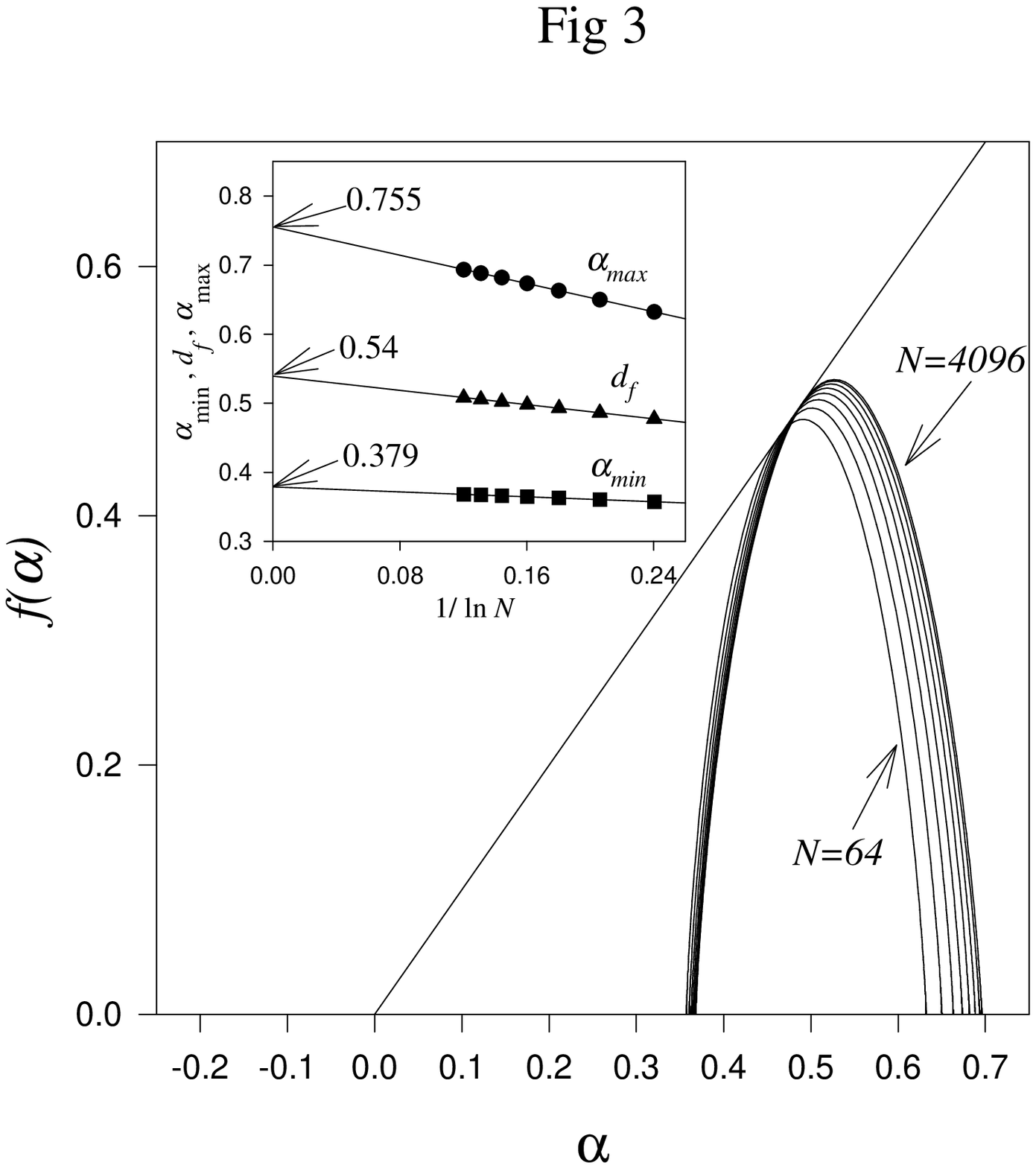}

\includegraphics{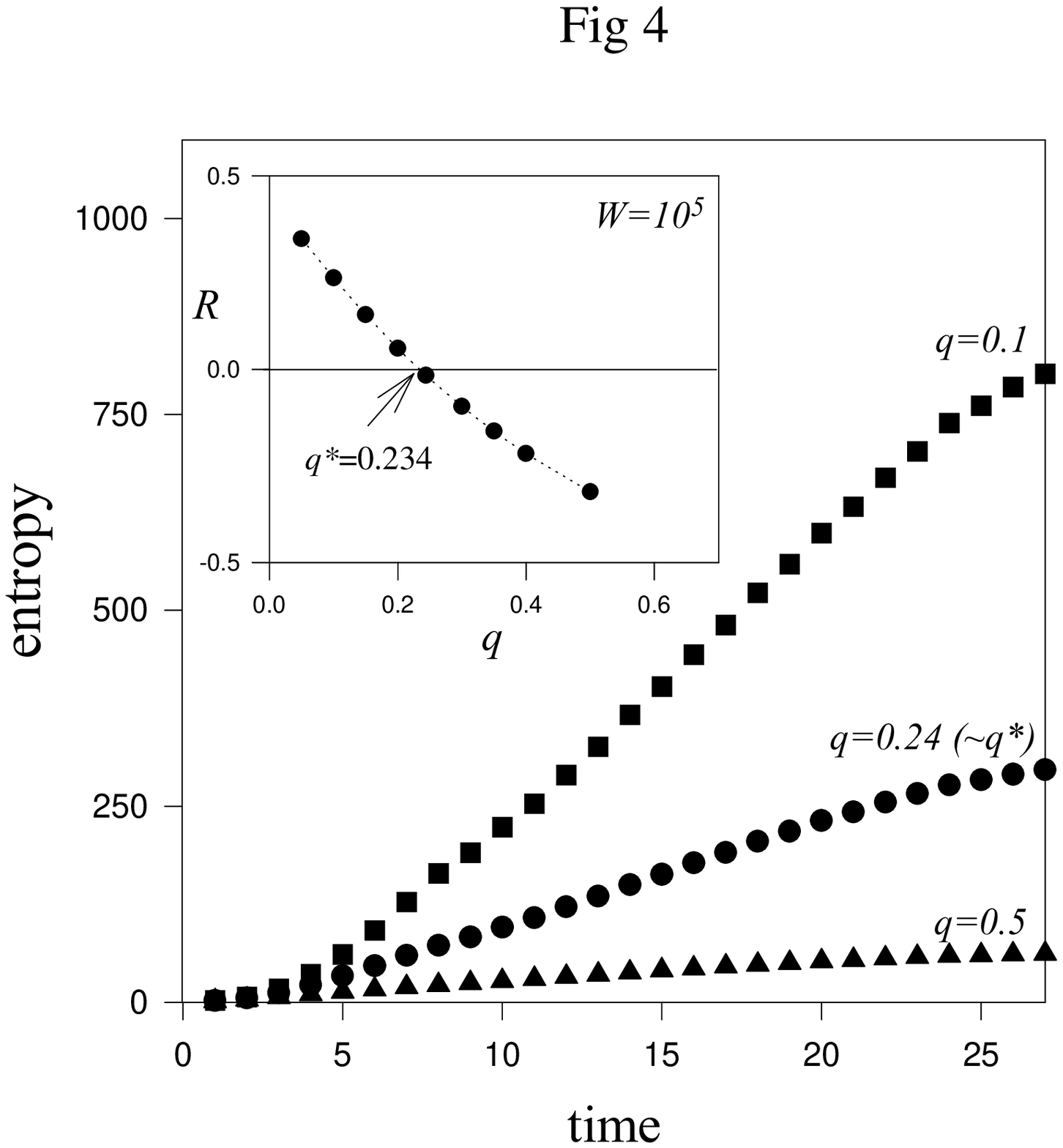}

\end{document}